\documentclass[a4paper]{jpconf}
\usepackage{graphicx}
\usepackage{amssymb}

\def\a{\alpha}
\def\b{\beta}

\def\p{\phi_a}
\def\ptil{\tilde\phi_a}

\begin{document}
\begin{flushright} OHSTPY-HEP-T-12-003
\end{flushright}

\title{Yukawa unification : MSSM at large $\tan \beta$}

\author{Stuart Raby}

\address{The Ohio State University, 191 W. Woodruff Ave, Columbus, OH 43210, USA}

\ead{raby@mps.ohio-state.edu}

\begin{abstract}
Talk given at PASCOS 2012,  Merida, Mexico describing work in progress in collaboration with Archana Anandakrishnan, Christopher Plumberg
and Akin Wingerter.
\end{abstract}

\section{Introduction}
In this talk we discuss SO(10) Yukawa unification and its ramifications for phenomenology.  The initial constraints come from fitting the
top, bottom and tau masses, requiring large $\tan\beta \sim 50$ and particular values for soft SUSY breaking parameters.  We will show that
this leads to a spectrum of squark and slepton masses with the first two families heavy while the third family is significantly lighter,
i.e. an inverted scalar mass hierarchy.  In addition, gauginos should be observable at the LHC.  We perform a global $\chi^2$ analysis,
fitting the recently observed `Higgs' with mass of order 125 GeV in addition to fermion masses and mixing angles and several flavor violating observables.

Fermion masses and mixing angles are manifestly hierarchical.   The simplest way to describe this hierarchy is with Yukawa matrices which
are also hierarchical.   Moreover the most natural way to obtain the hierarchy is in terms of effective higher dimension operators of the form
\begin{equation}  W \supset 16_3 \ 10 \ 16_3 + 16_3 \ 10 \ \frac{45}{M} \ 16_2 + \cdots .
\end{equation}
This version of SO(10) models has the nice features that it only requires small representations of SO(10),  has many predictions
and can, in principle, find an UV completion in string theory.  There are a long list of papers by authors such as Albright, Anderson, Babu,
Barr, Barbieri, Berezhiani, Blazek, Carena, Chang, Dermisek, Dimopoulos, Hall, Masiero, Murayama, Pati, Raby, Romanino, Rossi, Starkman,
Wagner, Wilczek, Wiesenfeldt, and Willenbrock which have followed this line of model building.

The only renormalizable term in $W$ is $\lambda \ 16_3 \ 10 \ 16_3$ which gives Yukawa coupling unification
\begin{equation}  \lambda = \lambda_t = \lambda_b = \lambda_\tau = \lambda_{\nu_\tau}  \end{equation} at $M_{GUT}$.
Note,  one CANNOT predict the top mass due to large SUSY threshold corrections to the bottom and tau masses, as shown in
\cite{down}.  These corrections are of the form
\begin{equation}  \delta m_b/m_b  \propto \frac{\alpha_3 \ \mu \ M_{\tilde g} \ \tan\beta}{m_{\tilde b}^2} +
\frac{\lambda_t^2 \ \mu \ A_t \ \tan\beta}{m_{\tilde t}^2} + {\rm log \ corrections} .
\end{equation} So instead  we use  Yukawa unification to predict the soft SUSY breaking masses!!  In order to fit the data,
we need \begin{equation} \delta m_b/m_b \sim - 2\% . \end{equation}  We take $\mu \ M_{\tilde g} > 0$, thus
we need $\mu \ A_t < 0$.  For a short list of references on this subject, see \cite{bdr,yukunif}.

Given the following GUT scale boundary conditions --  universal squark and slepton masses,  $m_{16}$,  universal cubic scalar parameter, $A_0$,
universal gaugino masses, $M_{1/2}$, and non-universal Higgs masses [NUHM] or `just so' Higgs splitting,  $m_{H_u}, \ m_{H_d}$  or
$m_{H_{u (d)}}^2 = m_{10}^2 [ 1 - (+) \Delta_{m_H}^2 ]$, we find that fitting the top, bottom and tau mass forces us into the region of SUSY breaking parameter
space with
\begin{equation}   A_0 \approx  - 2 m_{16},   \;\;  m_{10} \approx  \sqrt{2} \ m_{16}, \;\;  m_{16} > \ {\rm few \ TeV}, \;\;  \mu, M_{1/2} \ll m_{16}; \end{equation}
and, finally,  \begin{equation} \tan\beta \approx 50 . \end{equation}  In addition, radiative electroweak symmetry breaking requires
$\Delta_{m_H}^2 \approx 13\%$, with roughly half of this coming naturally from the renormalization group running of neutrino Yukawa couplings
from $M_G$ to $M_{N_\tau} \sim 10^{13}$ GeV \cite{bdr}.

It is very interesting that the above region in SUSY parameter space results in an inverted scalar mass hierarchy at the weak scale with the third family scalars
significantly lighter than the first two families \cite{bagger}.  This has the nice property of suppressing flavor changing
neutral current and CP violating processes.

\section{Heavy squarks and sleptons}

\begin{figure}[h]
\includegraphics[width=14pc]{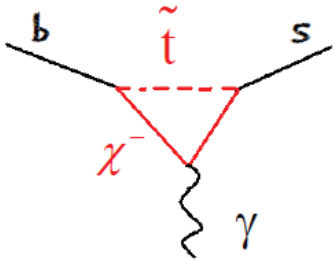}\hspace{2pc}%
\begin{minipage}[b]{14pc}\caption{\label{charginoloop}Dominant contribution to the process $b \rightarrow s \ \gamma$ in the MSSM.}
\end{minipage}
\end{figure}

Considering the theoretical and experimental results for the branching ratio $BR(B \rightarrow X_s \gamma)$, we argue that $m_{16} \ge 8$ TeV.   The experimental
value $BR(B \rightarrow X_s \gamma)_{\rm exp} = (3.55 \pm 0.26) \times 10^{-4}$, while the NNLO Standard Model theoretical value is
$BR(B \rightarrow X_s \gamma)_{\rm th} = (3.15 \pm 0.23) \times 10^{-4}$.  The amplitude for the process $B \rightarrow X_s \gamma$ is proportional to
the Wilson coefficient, $C_7$.   $C_7 = C_7^{SM} + C_7^{SUSY}$ and, in order to fit the data, we see that $C_7 \approx \pm C_7^{SM}$.  Thus
$C_7^{SUSY} \approx - 2 C_7^{SM} \; {\rm or} \; C_7^{SUSY} \approx 0$.
The dominant SUSY contribution to the branching ratio comes from a stop - chargino loop with
$C_7^{SUSY} \sim C_7^{\chi^+} \sim \frac{\mu \ A_t}{\tilde m^2} \ \tan\beta \times sign(C_7^{SM})$ (see Fig. \ref{charginoloop}).  Hence, in our
case  $C_7 \approx - C_7^{SM}$.

\begin{figure}[h]
\includegraphics[width=14pc]{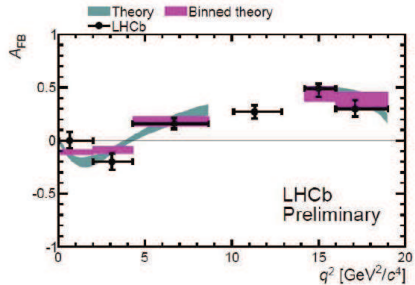}\hspace{2pc}%
\begin{minipage}[b]{14pc}\caption{\label{bkmumu}The forward-backward asymmetry for the process $B \rightarrow K^* \ \mu^+ \ \mu^-$ measured by LHCb.}
\end{minipage}
\end{figure}

Recent LHCb data on the $BR(B \rightarrow K^* \ \mu^+ \ \mu^-)$ now favors  $C_7 \approx + C_7^{SM}$ \cite{lhcb}.  See Fig. \ref{bkmumu}.
This tension between the processes  $b \rightarrow s \gamma$ and $b \rightarrow s \ \ell^+ \ \ell^-$ was already discussed by
Albrecht et al. \cite{albrecht}.  In order to be consistent with this
data one requires $C_7^{\chi^+} \approx 0$  or  $C_7 \approx C_7^{SM} + C_7^{SUSY} \approx + C_7^{SM}$ and therefore  $m_{16} \ge 8$ TeV.

\section{Light Higgs mass}

An approximate formula for the light Higgs mass is given by \cite{higgs}
\begin{equation} m_h^2 \approx M_Z^2 \ \cos^22\beta + \frac{3 g^2 m_t^4}{8 \pi^2 m_W^2} \left[\ln\left(\frac{M_{SUSY}^2}{m_t^2}\right) +
\frac{X_t^2}{M_{SUSY}^2} \left(1 - \frac{X_t^2}{12 M_{SUSY}^2}\right)\right] \end{equation}  where
$X_t = A_t - \mu/\tan\beta$.   The light Higgs mass is maximized as a function of $X_t$ for $X_t/M_{SUSY} = \sqrt{6}$, referred to as maximal
mixing.   Hence we see that for large values of $A_t$ and $M_{SUSY}$ it is quite easy to obtain a light Higgs mass of order 125 GeV.
\begin{figure}[h]
\includegraphics[width=14pc]{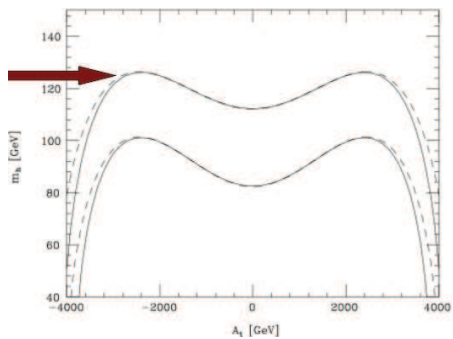}\hspace{2pc}%
\begin{minipage}[b]{14pc}\caption{\label{higgs} Light Higgs mass as a function of $A_t$ for two values of $\tan\beta$. The upper curve corresponds to
$\tan\beta \sim 50$. The arrow corresponds
to maximal mixing.}
\end{minipage}
\end{figure}

\section{$B_s \rightarrow \mu^+ \ \mu^-$}

In this section we argue that the light Higgs boson must be Standard Model-like.  To do this we show that the CP odd Higgs boson, $A$,
must have mass greater than $\sim$ 1 TeV and as a consequence this is also true for the CP even Higgs boson, $H$, and the charged Higgs bosons,  $H^\pm$, as well.
This is the well-known decoupling limit in which the light Higgs boson couples to matter just like the Standard Model Higgs.

Consider the branching ratio $BR(B_s \rightarrow \mu^+ \ \mu^-)$ which in the Standard Model is  $\sim 3 \times 10^{-9}$.   In the MSSM this receives
a contribution proportional to $\sim  \frac{\tan\beta^6}{m_A^4}$.   Recent experimental results give \cite{bsmm}
\begin{eqnarray}   CDF : &  1.8 (+ 1.8, - 0.9) \times 10^{-8} \; (95\% CL) & {\rm with} \; 7 fb^{-1} \\
LHCb \; {\rm bound} : &  < 4.5 \times 10^{-9} \; (95\% CL) &  {\rm with} \; 1 fb^{-1}.  \end{eqnarray}   These two results seem to be inconsistent, so we will choose
to try and satisfy the LHCb data which is also consistent with results from CMS.   Since we have $\tan\beta \sim 50$, our only choice is to take the
CP odd Higgs mass to be large with $m_A \ge 1$ TeV.   Hence the light Higgs is SM-like.

\section{3 Family Model}

The previous results depended solely on SO(10) Yukawa unification for the third family.   We now consider a complete three family SO(10) model for fermion masses
and mixing, including neutrinos \cite{3family,albrecht}.   The model also includes a
$D_3 \times [U(1) \times \mathbb{Z}_2 \times \mathbb{Z}_3]$ family symmetry which is necessary to
obtain a predictive theory of fermion masses by reducing the number of arbitrary parameters in the Yukawa matrices.
In the rest of this talk we will consider the new results due to the three family analysis.  We shall consider the superpotential generating the effective
fermion Yukawa couplings.  We then perform a global $\chi^2$ analysis, including precision electroweak data which now includes both neutral and charged
fermion masses and mixing angles.

\begin{figure}[h]
\includegraphics[width=20pc]{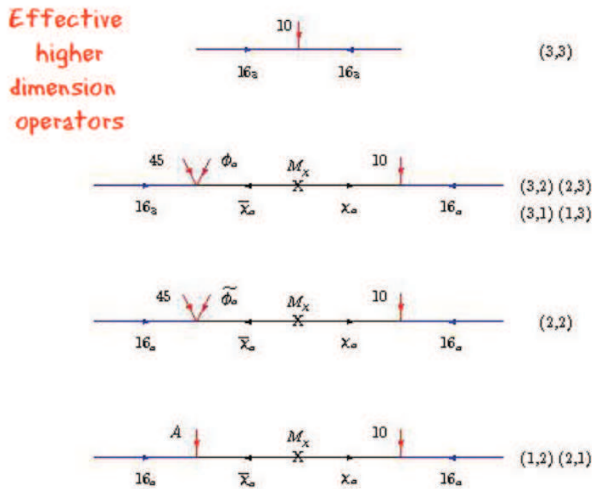}\hspace{2pc}%
\begin{minipage}[b]{20pc}\caption{\label{massoperators}The effective fermion mass operators obtained after integrating out the Froggatt-Nielsen massive states.}
\end{minipage}
\end{figure}

\begin{figure}[h]
\includegraphics[width=14pc]{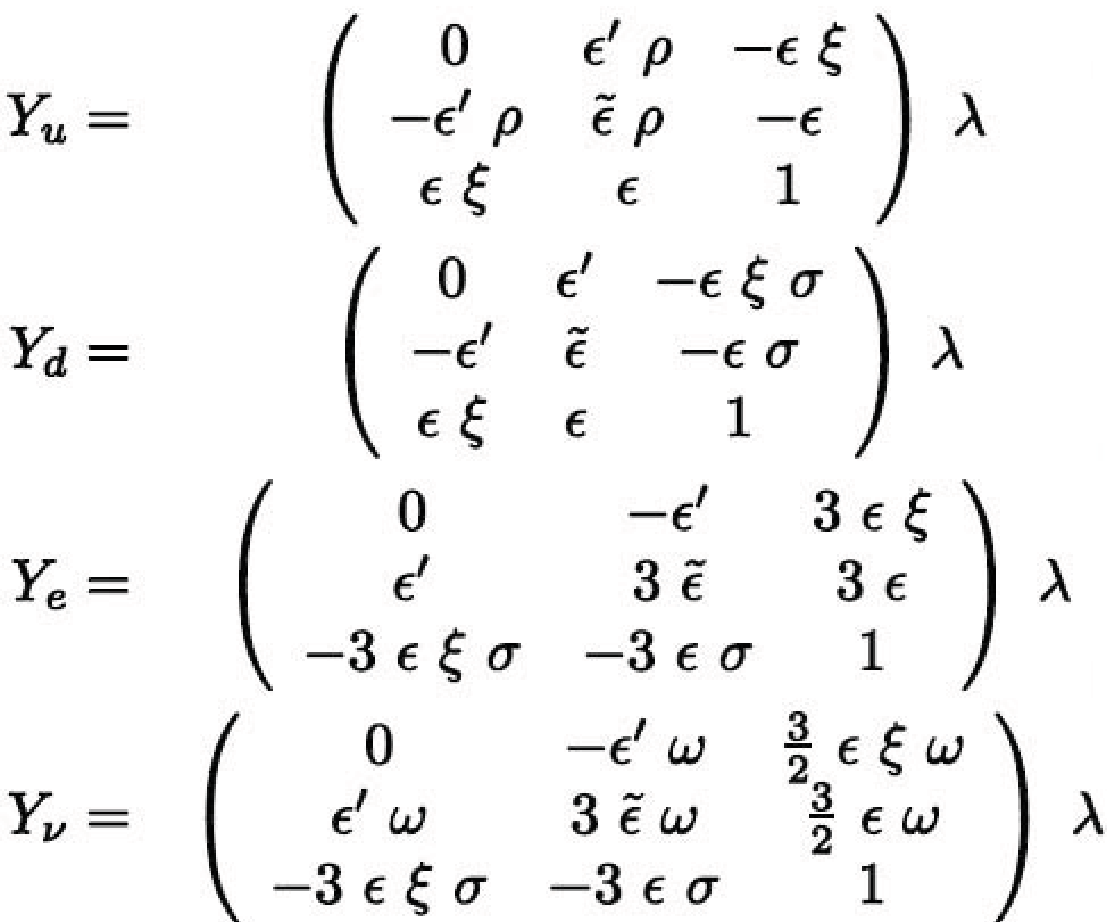}\hspace{2pc}%
\begin{minipage}[b]{14pc}\caption{\label{yukawas}The Yukawa matrices obtained from the effective fermion mass operators after taking into account the flavon VEVs.}
\end{minipage}
\end{figure}

The superspace potential for the charged fermion sector of this model is
given by:
\begin{eqnarray} W_{ch. fermions} = & 16_3 \ 10 \ 16_3 +  16_a \ 10 \ \chi_a & \label{Wchf}
\\ & +  \bar \chi_a \ ( M_{\chi} \ \chi_a + \ 45 \ \frac{\phi_a}{\hat M} \ 16_3 \ + \ 45 \ \frac{\tilde
\phi_a}{\hat M} \  16_a + {\bf A} \ 16_a ) & \nonumber
\end{eqnarray}
where $45$ is an $SO(10)$ adjoint field which is assumed to obtain
a VEV in the B -- L direction; and $M$ is a linear combination of an
$SO(10)$ singlet and adjoint. Its VEV $M_0 ( 1 + \a X + \b Y)$ gives mass
to Froggatt-Nielsen states. Here $X$ and $Y$ are elements of the Lie
algebra of $SO(10)$ with $X$ in the direction of the $U(1)$ which
commutes with $SU(5)$ and $Y$ the standard weak hypercharge; and $ \a $ ,
$ \b $ are arbitrary constants which are fit to the data.

\begin{equation}
\p, \nonumber
\quad \ptil, \nonumber
\quad A, \nonumber
\end{equation}
are $SO(10)$ singlet 'flavon' fields, and
\begin{equation}
\bar \chi_a, \nonumber
 \quad \chi_a
\nonumber
\end{equation}
are a pair of Froggatt-Nielsen states transforming as a $\overline {16}$ and $16$ under $SO(10)$.   The 'flavon' fields are {\em assumed} to obtain VEVs of the form
\begin{equation}  \langle \p \rangle = \left( \begin{array}{c} \phi_1 \\ \phi_2 \end{array} \right), \;\;
\langle \ptil \rangle = \left( \begin{array}{c} 0 \\ \tilde \phi_2 \end{array} \right). \end{equation}
After integrating out the Froggatt-Nielsen states one obtains the effective fermion mass operators in Fig. \ref{massoperators}.
We then obtain the Yukawa matrices for up and down quarks, charged leptons and neutrinos given in Fig. \ref{yukawas}.  These matrices contain 7
real parameters and 4 arbitrary phases.
Note,  the superpotential (Eqn. \ref{Wchf}) has many arbitrary parameters.   However, at the end of the day the effective Yukawa matrices have many fewer parameters.
This is good, because we then obtain a very predictive theory.   Also, the quark mass matrices accommodate the Georgi-Jarlskog mechanism, such that
$m_\mu/m_e \approx 9 m_s/m_d$.

We then add 3 real Majorana mass parameters for the neutrino see-saw mechanism. The anti-neutrinos get GUT scale masses by mixing with
three $SO(10)$ singlets $\{ N_a, \ a = 1,2;
\;\; N_3 \}$ transforming as a $D_3$ doublet and singlet respectively. The full superpotential is given by $W =
W_{ch. fermions} + W_{neutrino}$ with
\begin{eqnarray} \label{eq:WneutrinoD3} W_{neutrino} = & \overline{16} \left(\lambda_2 \ N_a \ 16_a \ + \ \lambda_3 \ N_3 \ 16_3 \right) & \\
& +  \;\; \frac{1}{2} \left(S_{a} \ N_a \ N_a \;\; + \;\; S_3 \ N_3 \ N_3\right).   & \nonumber
\end{eqnarray}
We assume $\overline{16}$ obtains a VEV, $v_{16}$, in the right-handed neutrino direction, and $\langle S_{a}
\rangle = M_a$ for $a = 1,2$ and $\langle S_3 \rangle = M_3$.  The effective neutrino mass terms are given by
\begin{equation} W =  \nu \ m_\nu \ \bar \nu + \bar \nu \ V \ N +
\frac{1}{2} \ N \ M_N \ N \end{equation} with
\begin{equation} V = v_{16} \ \left(
\begin{array}{ccc} 0 &  \lambda_2 & 0 \\
\lambda_2 & 0 & 0 \\ 0 & 0 &  \lambda_3 \end{array} \right), \; M_N = diag( M_1,\ M_2,\ M_3)  \end{equation}  all assumed
to be real.  Finally, upon integrating out the heavy Majorana neutrinos we obtain the $3 \times 3$ Majorana mass matrix for the light
neutrinos in the lepton flavor basis given by
\begin{equation}
{\cal M} =   U_e^T \ m_\nu  \ M_R^{-1} \ m_\nu^T  \ U_e ,
\end{equation}
where the effective right-handed neutrino Majorana mass matrix is given by:
\begin{equation}
M_R =  V \ M_N^{-1}  \ V^T  \  \equiv \  {\rm diag} ( M_{R_1}, M_{R_2}, M_{R_3} ),
\end{equation}
with \begin{eqnarray} M_{R_1} = (\lambda_2 \ v_{16})^2/M_2, \quad  M_{R_2} = (\lambda_2 \ v_{16})^2/M_1, \quad  M_{R_3} =
(\lambda_3 \ v_{16})^2/M_3 . \label{eq:rhmass} \end{eqnarray}

\section{Global $\chi^2$ analysis}

Just in the fermion mass sector we can see that the theory is very predictive.   We have 15 charged fermion and 5 neutrino low energy observables given
in terms of 11 arbitrary Yukawa parameters and 3 Majorana mass parameters.  Hence there are 6 degrees of freedom in this sector of the theory.   However
in order to include the complete MSSM sector we perform the global $\chi^2$ analysis with 24 arbitrary parameters at the GUT scale given in Table \ref{paras}.
Note, this is to be compared to the 27 arbitrary parameters in the Standard Model or the 32 parameters in the CMSSM.
\begin{table}[h]
\caption{\label{paras}Parameters entering the global $\chi^2$ analysis.}
\begin{center}
\lineup
\begin{tabular}{*{3}{l}}
\br
$\0\0Sector$&\#&$Parameters$\cr
\mr
\0$gauge$&$3$  &$\alpha_G, \ M_G, \ \epsilon_3$\cr
\0$SUSY \ (GUT scale)$&$5$  &$m_{16}, \ M_{1/2}, \ A_0, \ m_{H_u}, \ m_{H_d}$\cr
\0$textures$&$11$  &$\lambda, \ \epsilon, \ \epsilon', \ \rho, \ \sigma, \ \tilde \epsilon, \ \xi$\cr
\0$neutrino$&$3$  &$M_{R_1}, \ M_{R_2}, \ M_{R_3}$\cr
\0$SUSY \ (EW scale)$&$2$  &$\mu, \ \tan\beta$\cr
\br
\end{tabular}
\end{center}
\end{table}

\begin{figure}[h]
\includegraphics[width=30pc]{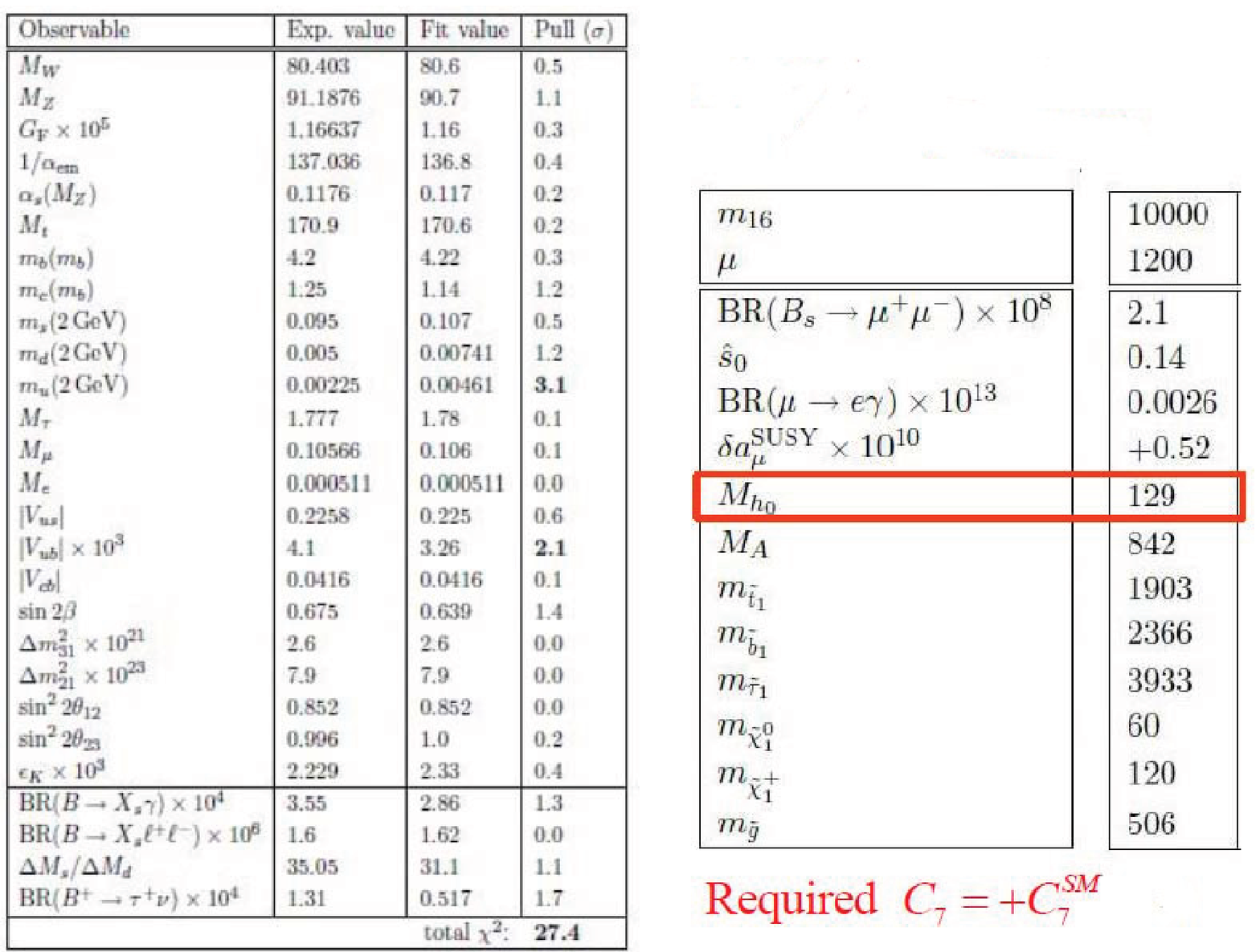}\hspace{2pc}%
\begin{minipage}[b]{30pc}\caption{\label{albrecht}The results obtained by Albrecht et al. for $m_{16} = 10$ TeV.}
\end{minipage}
\end{figure}

In 2007,  Albrecht et al. \cite{albrecht} performed a global $\chi^2$ analysis of this theory.  Two of the tables from their paper are exhibited in
Fig. \ref{albrecht}.
This analysis included 27 low energy observables and a reasonable fit to the data was only found for $m_{16} = 10$ TeV.   Note, the Higgs mass was predicted
to be 129 GeV.
There is now a discovery by CMS and ATLAS of a boson with mass of order 125 GeV.   This may well be the Higgs boson with Standard Model couplings, but
further data is needed before such a conclusion can be confirmed.  In the mean time we shall assume it is the lightest Higgs boson of the MSSM.  In this
case, we must conclude (based on our analysis) that it must look very much like the Standard Model Higgs boson, since as can be seen from Fig. \ref{albrecht}, the
CP odd Higgs mass, $m_A = 842$ GeV, so the theory is in the Higgs decoupling limit.

In this work we have decided to extend the analysis of Albrecht et al. to values of $m_{16} \ge 10$ TeV,  including more low energy observables such as the light
Higgs mass,  the neutrino mixing angle $\theta_{13}$ and lower bounds on the gluino and squark masses coming from recent data.  This is work in progress in
collaboration with Archana Anandakrishnan, Christopher Plumberg and Akin Wingerter, along with significant help from Radovan Dermisek and Aditi Raval.
We perform a three family global $\chi^2$ analysis.  We are using the code developed by Radovan Dermisek to renormalize the parameters in the theory from the
GUT scale to the weak scale,  perform electroweak symmetry breaking and calculate squark, slepton, gaugino masses, as well as quark and lepton masses and mixing angles.
We also use the Higgs code of Pietro Slavich (suitably revised for our particular scalar spectrum) to calculate the light Higgs mass and SUSY\_Flavor\_v2.0 \cite{SFv2}
to evaluate flavor violating B decays.

There are 24 arbitrary parameters defined mostly at the GUT scale and run down to the weak scale where the $\chi^2$ function is evaluated.  However
the value of $m_{16}$ has been kept fixed in our analysis, so that we can see the dependence of $\chi^2$ on this input parameter.   Thus with
23 arbitrary parameters we fit 37 observables, giving 14 degrees of freedom.  The $\chi^2$ function has been minimized using the CERN package, MINUIT.
This work is on going so we just present the complete results
for one value of $m_{16} = 20$ TeV (see Table \ref{20tev}).   The fit is quite good with $\chi^2/d.o.f. = 1.8$.   However, note that we have not
taken into account correlations in the data, so we will just use $\chi^2$ as a indicator of the rough goodness of the fit.
\begin{table}
\caption{\label{20tev}Result of global $\chi^2$ analysis for $m_{16}$ fixed at 20 TeV.}
\begin{tabular}{|l|l|l|l|l|}
\hline
Observable  &  Fit value  &  Exp. value  &  Pull & Sigma  \\
\hline
$M_Z$ &              91.1876         &  91.1876         &  0.0000          &  0.0912          \\
$M_W$ &              80.7955         &  80.4360         &  0.8938          &  0.4022          \\
$1/\alpha_{em}$ &    136.5851        &  137.0360        &  0.6580          &  0.6852          \\
$G_{\mu} \times 10^5$ & 1.1766          &  1.1664          &  0.8805          &  0.0117          \\
$\alpha_3$ &         0.1184          &  0.1184          &  0.0642          &  0.0007          \\
\hline
$M_t$ &              173.2432        &  172.9000        &  0.3177          &  1.0800          \\
$m_b(m_b)$ &         4.3155          &  4.1900          &  0.6973          &  0.1800          \\
$m_{\tau}$ &         1.7766          &  1.7770          &  0.0484          &  0.0089          \\
\hline
$M_b-M_c$ &          3.4285          &  3.4300          &  0.0300          &  0.0500          \\
$m_c(m_c)$ &         1.1723          &  1.2900          &  1.0697          &  0.1100          \\
$m_s$ &              0.0969          &  0.1000          &  0.1039          &  0.0300          \\
$m_d/m_s$  &         0.0693          &  0.0521          &  2.5690          &  0.0067          \\
$1/Q^2$ &            0.0019          &  0.0019          &  0.3475          &  0.0001          \\
$M_{\mu}$ &          0.1056          &  0.1057          &  0.0386          &  0.0005          \\
$M_e \times 10^4$ &  5.1125          &  5.1100          &  0.0978          &  0.0255          \\
\hline
$|V_{us}|$ &         0.2250          &  0.2252          &  0.2407          &  0.0009          \\
$|V_{cb}|$ &         0.0412          &  0.0406          &  0.4702          &  0.0013          \\
$|V_{ub}| \times 10^3$ & 3.2245          &  3.8900          &  1.5124          &  0.4400          \\
$|V_{td}| \times 10^3$ & 8.9128          &  8.4000          &  0.8547          &  0.6000          \\
$|V_{ts}|$ &         0.0404          &  0.0429          &  0.9750          &  0.0026          \\
${\rm sin} 2\beta$ & 0.6370          &  0.6730          &  1.5642          &  0.0230          \\
\hline
$\epsilon_K$ &       0.0021          &  0.0022          &  0.3631          &  0.0002          \\
$\Delta M_{Bs}/\Delta M_{Bd}$ & 31.7861         &  35.0000         &  0.4439          &  7.2400          \\
$\Delta M_{Bd} \times 10^{13}$ & 4.2657          &  3.3370          &  1.3944          &  0.6660          \\
\hline
$m^2_{21} \times 10^5$ &   7.5243          &  7.5400          &  0.0340          &  0.4600          \\
$m^2_{31} \times 10^3$ &  2.4708          &  2.4600          &  0.0637          &  0.1700          \\
${\rm sin}^2 \theta_{12}$ & 0.2984          &  0.3070          &  0.2445          &  0.0350          \\
${\rm sin}^2 \theta_{23}$ & 0.4390          &  0.3980          &  0.5323          &  0.0770          \\
${\rm sin}^2 \theta_{13}$ & 0.0132          &  0.0245          &  1.7062          &  0.0066          \\
\hline
$M_h$ &              124.80          &  125.30          &  0.1664          &  3.0000          \\
\hline
$BR(B \rightarrow X_s \gamma) \times 10^4$ & 3.4988          &  3.5500          &  0.0368          &  1.3900          \\
$BR(B_s \rightarrow \mu^+ \mu^-) \times 10^9$ & 4.2909          &  4.5000          &  0.0000          &  1.0000          \\
$BR(B_d \rightarrow \mu^+ \mu^-) \times 10^{10}$ & 1.3112          &  14.0000         &  0.0000          &  10.0000         \\
$BR(B \rightarrow \tau \nu) \times 10^5$ & 6.8017          &  16.5000         &  1.0428          &  9.3000          \\
$BR(B \rightarrow K^*\mu^+ \mu^-)$(low) $\times 10^8$& 5.0639          &  4.2000          &  0.3199          &  2.7000          \\
$BR(B \rightarrow K^*\mu^+ \mu^-)$(high) $\times 10^8$ & 7.5709          &  6.3000          &  0.4707          &  2.7000          \\
$q_0^2(B \rightarrow K^* \mu^+ \mu^-)$ & 4.6274          &  4.9000          &  0.2097          &  1.3000          \\
\hline
\multicolumn{3}{|l}{Total $\chi^2$}  &  \textbf{24.2}&  \\\hline
\end{tabular}
\end{table}
In Table \ref{output} we see that the value of $\chi^2$ decreases as $m_{16}$ increases,  but our analysis shows that the
$m_{16} \sim 20$ TeV minimizes $\chi^2$, i.e. we have found that $\chi^2$ slowly increases for $m_{16} > 20$ TeV.
\begin{table}[h]
\caption{\label{output}Results for the global $\chi^2$ analysis with 3 different values of $m_{16} = 10, 15, 20$ TeV.}
\begin{center}
\begin{tabular}{|l|l|l|l|}
\hline
$m_{16}  $  &  10 TeV  &  15 TeV &  20 TeV \\
$A_0$  &  -20 TeV & -30 TeV & -41 TeV     \\
$\mu  $  & 517 & 900 & 766      \\
$M_{1/2}$ &  270 & 200 & 169       \\
\hline
$\chi^2$ & 32.12 & 25.87 & 24.51    \\
\hline
$M_A$ &  2357 & 2743 & 5970     \\
$m_{\tilde t_1}$ & 1699 & 2952 & 3738     \\
$m_{\tilde b_1}$  & 2087 & 3481 & 3702     \\
$m_{\tilde \tau_1}$  & 3919 & 5932 & 7409       \\
$m_{\tilde\chi^0_1}$   & 145 & 133 & 136        \\
$m_{\tilde\chi^+_1}$  & 277 & 267 & 279        \\
$m_{\tilde g}$   & 925 & 850 & 850      \\
\hline
\end{tabular}
\end{center}
\end{table}

\section{Conclusion and plans for the future}

We have presented an analysis of a theory satisfying Yukawa unification and large $\tan\beta$.  The results are encouraging but as usual there are
some good and some bad features.

Some of the good features are -
\begin{itemize}
\item  gauge coupling unification is satisfied
\item  the Higgs mass is of order 125 GeV and Standard Model like
\item  there is an inverted scalar mass hierarchy, with the following consequence at the LHC -
\item  gluinos predominantly decay via $\tilde g \rightarrow  t \ \bar t \ \tilde \chi^0, \; t \ \bar b \ \tilde \chi^-, \;  b \ \bar b \ \tilde \chi^0$, etc.
\end{itemize}

And bad features -
\begin{itemize}
\item  fine-tuned by at least 0.1 \%
\item  and gluinos are light.
\end{itemize}

Work in progress -
\begin{itemize}
\item We would like to understand why $\chi^2$ increases for values of $m_{16} > 20$ TeV or is there an
upper bound on $m_{16}$?
\item The gaugino masses are small.   We want to evaluate the $\chi^2$ dependence as a function of the gluino mass or, in other words,
ascertain the sensitivity of $\chi^2$ on varying the gaugino masses?
\item Finally, we would like to consider alternative sets of boundary conditions at the GUT
scale which may be consistent with Yukawa unification.  In particular, we will consider the so-called ``DR3" scheme and non-universal gaugino masses as may
be obtained with (a) non-trivial gauge kinetic functions or (b) ``mirage mediation."
\end{itemize}

In the ``DR3" scheme \cite{dr3} we have SO(10) boundary conditions at $M_{GUT}$, however the third family scalar masses are taken to
be smaller than the first two, i.e.  $m_{16}^{(3)} \le m_{16}^{(1,2)}$.   In addition,  Higgs doublet mass splitting is accomplished with a U(1)$_X$ D-term, $D_X$,
where U(1)$_X$ is the is found in SO(10)/[SU(5) $\times$ U(1)$_X$].  Hence,  $m_i^2 = Q^i_X \ D_X + (m_i^0)^2$  and the charges, $Q^i_X$, are given in
Table \ref{Xcharge}.
\begin{table}[h]
\caption{\label{Xcharge}U(1)$_X$ charge and SO(10) invariant mass.}
\begin{center}
\begin{tabular}{lllllll|ll}
\br
  &$Q$ & $U$ & $D$ & $L$ & $E$ & $N$&$H_u$ & $H_d$\\
\mr
 $Q^i_X$&$1$ & $1$ & $-3$ & $-3$ & $1$ & $5$  & $-2$ & $2$\\
\mr
 $m_i^0$& &  &$m_{16}$ & & & & $\;\;\;\;  m_{10}$ & \\
\br
\end{tabular}
\end{center}
\end{table}

One possible scheme for obtaining non-universal gaugino masses was contemplated in \cite{badziak} using
a non-trivial gauge kinetic function with a chiral 54 dimensional representation of SO(10) obtaining an F-term VEV, but NO scalar VEV.  This gives
gaugino masses in the ratio $M_3 : M_2 : M_1 = 1 : -3/2 : -1/2$.  In addition they take $\mu < 0$ and U(1)$_X$ D-term splitting of squark, slepton and
Higgs masses.  In the end they obtain a totally different spectrum than with either ``Just So" Higgs splitting or ``DR3" splitting.

Finally, it has been shown that non-universal gaugino masses can be obtained with ``mirage mediation" boundary conditions at $M_{GUT}$ \cite{mirage}.
We would have gaugino masses given by
\begin{equation}
M_i(\mu)/g^2_i(\mu) = \left\{ 1 + \log(M_{Pl}/m_{3/2}) \frac{\alpha_{GUT} \ b_i}{4 \pi} \alpha \right\} M_0/g_{GUT}^2
\end{equation}
where $\alpha$ is an arbitrary parameter fixing the ratio of the contributions of gravity vs anomaly mediation.   In addition we would
take $\mu < 0$ and U(1)$_X$ D-term splitting for scalars.

\section*{Acknowledgements}
I would like to thank the organizers of PASCOS 2012 for organizing such a wonderful conference.  In addition,  I am very thankful to Radovan Dermisek for
letting us use his 3 family code which has been great and for the aid he gave in helping us implement the code along with his student, Aditi Raval.
Finally I want to thank Pietro Slavich for
sending us a copy of his code for calculating the light Higgs mass.  This work has been partially supported by DOE grant - DOE/ER/01545-895.


\section*{References}
\begin{thebibliography}{9}
\bibitem{down} L.J.~Hall, R.~Rattazzi and U.~Sarid, Phys.\ Rev.\ D {bf 50}, 7048 (1994); M.~Carena, M.~Olechowski, S.~Pokorski and C.~Wagner, Nucl.\ Phys.\ B
{\bf 426}, 269 (1994); T.~Blazek, S.~Raby and S.~Pokorski,  Phys.\ Rev.\ D{\bf 52}, 4151 (1995).
\bibitem{bdr} T.~Blazek, R.~Dermisek and S.~Raby,
Phys.\ Rev.\ Lett.\  {\bf 88}, 111804 (2002) [arXiv:hep-ph/0107097];
Phys.\ Rev.\ D {\bf 65}, 115004 (2002) [arXiv:hep-ph/0201081].
\bibitem{yukunif} H.~Baer and J.~Ferrandis,  Phys.\ Rev.\ Lett. {\bf 87}, 211803 (2001);
D.~Auto, H.~Baer, C.~Balazs, A.~Belyaev, J.~Ferrandis and X.~Tata,
JHEP 0306:023 (2003); K.~Tobe and J.~D.~Wells,
Nucl.\ Phys.\ B {\bf 663}, 123  (2003);
R.~Dermisek, S.~Raby, L.~Roszkowski and R.~Ruiz de Austri, JHEP 0304:037 (2003); ibid., JHEP 0509:029  (2005);
H.~Baer, S.~Kraml, S.~Sekmen and H.~Summy, JHEP 0803:056  (2008); ibid., JHEP 0810:079  (2008).
\bibitem{bagger} J.~A.~Bagger, J.~L.~Feng, N.~Polonsky and R.~J.~Zhang,
Phys.\ Lett.\ B {\bf 473}, 264 (2000).
\bibitem{albrecht}  M.~Albrecht, W.~Altmannshofer, A.J.~Buras, D.~Guadagnoli and D.M.~Straub, JHEP 0710:055 (2007).
\bibitem{lhcb} LHCb-Talk-2012-040 by Chris Parkinson, Imperial College (2012); LHCb arXiv:1208.3355 [hep-ex].
\bibitem{higgs}  M.~Carena, M.~Quiros and C.~Wagner, Nucl.\ Phys.\ B {\bf 461}, 407 (1996).
\bibitem{bsmm} R.~Aaij et al. [LHCb Collaboration], arXiv:1203.4493 [hep-ex]; T.~Aaltonen et al. [CDF Collaboration], Phys.\ Rev.\ Lett. {\bf 107}, 239903 (2011)
[Phys.\ Rev.\ Lett. {\bf 107}, 191801 (2011)] [arXiv:1107.2304].
\bibitem{3family} R.~Dermisek and S.~Raby, Phys.\ Lett.\ B {bf 622}, 327 (2005); R.~Dermisek, M.~Harada and S.~Raby, Phys.\ Rev.\ D {\bf 74}, 035011 (2006).
\bibitem{SFv2} SUSY\_Flavor\_v2, arXiv:submit/0440384 [hep-ph].
\bibitem{dr3} H.~Baer, S.~Kraml and S.~Sekman, [arXiv:0908.0134].
\bibitem{badziak} M.~Badziak, M.~Olechowski and S.~Pokorski, arXiv:1107.2764 [hep-ph].
\bibitem{mirage} O.~Lebedev, H.P.~Nilles and M.~Ratz, Phys.\ Lett.\ B {\bf 636}, 126 (2006);  K.~Choi and K.S.~Jeong, JHEP 0701, 103 (2007);
K.~Choi and H.P.~Nilles, arXiv:hep-ph/0702146.

\end{thebibliography}
\end{document}